\documentstyle[aps,psfig]{revtex}
\begin{document}

\title{Inflation: flow, fixed points and observables to arbitrary order in slow
 roll}
\author{William H.\ Kinney\thanks{Electronic address: {\tt
kinney@physics.columbia.edu}}} \address{
Institute for Strings, Cosmology and Astroparticle Physics\\
Columbia University\\
550 W. 120th St., New York, NY 10027
}
\date{May 1, 2003}
\maketitle

\begin{abstract}
I generalize the inflationary flow equations of Hoffman and Turner to arbitrary
order in slow roll. This makes it possible to study the predictions of
slow roll inflation in the full observable parameter space of tensor/scalar
ratio $r$, spectral index $n$, and running $d n / d \ln k$.  It also becomes
possible to identify exact fixed points in the parameter flow. I numerically
evaluate the flow equations to fifth order in slow roll for a set of randomly
chosen initial conditions and find that the models cluster strongly in the
observable parameter space, indicating a ``generic'' set of predictions for slow
roll inflation. I comment briefly on the the interesting proposed correspondence
between flow in inflationary parameter space and renormalization group flow in a
boundary conformal field theory.
\end{abstract}


\section{Introduction}

Inflationary cosmology\cite{guth81,linde82,albrecht82} has become the dominant
paradigm for describing the very early universe. Over the past twenty years,
inflationary model building has been a prolific enterprise\cite{lyth99}.
Concurrently, cosmological observations have improved to the point that it is
beginning to be possible to rule out models of
inflation\cite{kinney01,hannestad02}. Future observations, particularly the
MAP\cite{MAP} and Planck\cite{Planck} Cosmic Microwave Background (CMB)
satellites, promise to dramatically improve the situation in the near
future\cite{dodelson97,kinney98}. The key observational parameters for
distinguishing among inflation models are the tensor/scalar ratio $r$, the
scalar spectral index $n$, and the ``running'' of the spectral index, $d n / d
\ln k$, since different inflation models predict different values for these
parameters.

It is desirable, however, to gain some insight into what the {\em generic}
predictions of inflation are without having to work within the context of some
particular model. The standard lore of a small tensor/scalar ratio and nearly
scale-invariant power spectrum is insufficient now that precision measurements
of the CMB and large-scale structure are becoming a reality. Hoffman and Turner
have proposed the method of inflationary ``flow'' to gain generic insight into
the behavior of inflation models\cite{hoffman00}. The flow equations relate the
time derivatives of the slow roll parameters to other, higher order slow roll
parameters. With a suitable choice of truncation, this makes it possible to
study the dynamics of inflation models without having to specify a particular
potential for the field driving inflation. In this paper we generalize the
method from the lowest-order analysis of Hoffman and Turner and derive a simple
set of flow equations which can be evaluated to arbitrarily high order, and
which are in fact {\em exact} in the limit of infinite order in slow roll. We
perform a numerical integration of $10^{5}$ inflation models to fifth order in
slow roll, and plot their predictions in the observable parameter space
$(r,n,dn/d\ln k)$. The predictions of the models cluster strongly in the
observable parameter space, in fact even more strongly than was suggested by
Hoffman and Turner. (However the qualitative character of their analysis is
preserved at higher order in slow roll.) We emphasize that in this paper we
limit ourselves to inflation driven by a single scalar field $\phi$. The
case of multiple-field inflation is in general much more complex. 

This idea of flow in the inflationary parameter space has taken on different
significance with recent ideas arising from the ``holographic'' correspondence
between de Sitter space and boundary conformal field theories proposed by
Strominger\cite{strominger01a}. Particularly interesting are efforts to
interpret flow in the space of slow roll parameters as renormalization group
flow in a boundary conformal field theory\cite{larsen02}. This raises the
possibility that understanding the evolution of inflationary parameters is
important not just for phenomenology, but for fundamental reasons as well.

The paper is organized as follows: Section \ref{sechjreview} briefly reviews the
very powerful Hamilton-Jacobi formalism for inflation. Section
\ref{secperturbations} discusses the generation of fluctuations in inflation
and the relationship between the slow roll parameters and the observables in
various exact and approximate solutions of the inflationary equations of
motion. The hierarchy of flow equations is derived in Section
\ref{secsrhierarchy}.  Section \ref{secfixedpoints} discusses the fixed points
in the slow roll parameter space. Section \ref{secevaluatingtheflowequations}
discusses the details of the numerical solution. Section \ref{secconclusions}
presents conclusions.

\section{Inflation and The Hamilton-Jacobi Formalism}
\label{sechjreview}

The dominant component of an inflationary cosmology is a spatially homogeneous
scalar field $\phi$ (the {\em inflaton}) with potential $V\left(\phi\right)$
and equation of motion \begin{equation}
\ddot \phi + 3 H \dot \phi + V'\left(\phi\right) = 0,\label{eqequationofmotion}
\end{equation}
where $H \equiv (\dot a / a)$ is the Hubble parameter, and the Einstein field
 equations for the evolution of a flat background metric 
\begin{equation}
ds^2 = dt^2 - a^2\left(t\right) d {\bf x}^2
\end{equation}
can be written as
\begin{equation}
H^2 = \left({\dot a \over a}\right)^2 = {8 \pi \over 3 m_{\rm Pl}^2}
 \left[V\left(\phi\right) + {1 \over 2}
 \dot\phi^2\right],\label{eqbackgroundequation1}
\end{equation}
and
\begin{equation}
\left({\ddot a \over a}\right) = {8  \pi \over 3 m_{\rm Pl}^2}
 \left[V\left(\phi\right) - \dot\phi^2\right].\label{eqbackgroundequation2}
\end{equation}
Here $m_{\rm Pl} = G^{-1/2} \simeq 10^{19}\,{\rm GeV}$ is the Planck mass. These
background equations, along with the equation of motion
(\ref{eqequationofmotion}), form a coupled set of differential equations
describing the evolution of the universe. In the limit that $\dot\phi = 0$, the
expansion of the universe is of the de Sitter form, with the scale factor
increasing exponentially in time: \begin{eqnarray}
&&H = \sqrt{\left({8 \pi \over 3 m_{\rm Pl}^2}\right) V\left(\phi\right)} = {\rm
 const},\cr
&&a \propto e^{H t}.
\end{eqnarray}
In general, the Hubble parameter $H$ will not be exactly constant, but will vary
as the field $\phi$ evolves along the potential $V\left(\phi\right)$. A
convenient approach to the more general case is to express the Hubble parameter
directly as a function of the field $\phi$ instead of as a function of time, $H
= H\left(\phi\right)$. This is consistent as long as $\phi$ is monotonic in
time. The equations of motion for the field and background are  given
by\cite{grishchuk88,muslimov90,salopek90,lidsey95}: \begin{eqnarray} &&\dot\phi
= -{m_{\rm Pl}^2 \over 4 \pi} H'\left(\phi\right),\cr
&&\left[H'\left(\phi\right)\right]^2 - {12 \pi \over m_{\rm Pl}^2}
H^2\left(\phi\right) = - {32 \pi^2 \over m_{\rm Pl}^4}
V\left(\phi\right).\label{eqbasichjequations} \end{eqnarray}
These equations are completely equivalent to the second-order equation of motion
({\ref{eqequationofmotion}). The second of these is referred to as the {\it
Hamilton-Jacobi} equation, and can be written in the useful form
\begin{equation} H^2\left(\phi\right) \left[1 - {1\over 3}
\epsilon\left(\phi\right)\right] =  \left({8 \pi \over 3 m_{\rm Pl}^2}\right)
 V\left(\phi\right),\label{eqhubblehamiltonjacobi}
\end{equation}
where the parameter $\epsilon$ is defined as
\begin{equation}
\epsilon \equiv {m_{\rm Pl}^2 \over 4 \pi} \left({H'\left(\phi\right) \over
 H\left(\phi\right)}\right)^2.\label{eqdefofepsilon}
\end{equation}
The physical meaning of the parameter $\epsilon$ can be seen by expressing Eq. 
 (\ref{eqbackgroundequation2}) as
\begin{equation}
\left({\ddot a \over a}\right) = H^2 \left(\phi\right) \left[1 -
 \epsilon\left(\phi\right)\right],
\end{equation}
so that the condition for inflation $(\ddot a / a) > 0$ is given by $\epsilon <
 1$. The evolution of the scale factor is given by the general expression
\begin{equation}
a \propto \exp\left[\int_{t_0}^{t}{H\,dt}\right],
\end{equation}
where the number of e-folds $N$ is defined to be
\begin{equation}
N \equiv \int_{t}^{t_e}{H\,dt} = \int_{\phi}^{\phi_e}{{H \over \dot\phi}\,d\phi}
 =  {2 \sqrt{\pi} \over m_{\rm Pl}} \int_{\phi_e}^{\phi}{d\phi \over
 \sqrt{\epsilon\left(\phi\right)}}.\label{eqdefofN}
\end{equation}
Here we take $t_e$ and $\phi_e$ to be the time and field value at end of
inflation. Therefore $N$ increases as one goes {\em backward} in time, $d t > 0
\Rightarrow d N < 0$. These expressions are exact, and do not depend on any
assumption of slow roll. It is important to note the sign convention for
$\sqrt{\epsilon}$, which we define to have the same sign as
$H'\left(\phi\right)$: \begin{equation}
\sqrt{\epsilon} \equiv + {m_{\rm PL} \over 2 \sqrt{\pi}} {H' \over H}.
\end{equation}
In the next section we briefly discuss the generation of perturbations in
inflation from the point of view of different exact and approximate solutions.

\section{Cosmological perturbations: scalar and tensor power spectra}
\label{secperturbations}

Cosmological density and gravitational wave perturbations in the inflationary
scenario arise as quantum fluctuations which are ``redshifted'' to long
wavelength by the rapid cosmological
expansion\cite{hawking82,starobinsky82,guth82,bardeen83}. The power spectrum of
density perturbations is given by\cite{mukhanov92} \begin{equation}
P_{\cal R}^{1/2}\left(k\right) = \sqrt{k^3 \over 2 \pi^2} \left|{u_k \over
 z}\right|,
\end{equation}
where $k$ is a comoving wave number, and the mode function $u_k$ satisfies the
differential equation\cite{stewart93,mukhanov85,mukhanov88}
\begin{equation}
{d^2 u_k \over d\tau^2} + \left(k^2 - {1 \over z} {d^2 z \over d\tau^2} \right)
 u_k = 0,\label{eqexactmodeequation}
\end{equation}
where $\tau$ is the conformal time, $ds^2 = a^2\left(\tau\right) \left(d \tau^2
 - d {\bf x}^2 \right)$, and the quantity $z$ is defined as
\begin{equation}
z \equiv {2 \sqrt{\pi} \over m_{\rm Pl}} \left({a \dot\phi \over H}\right) = - a
 \sqrt{\epsilon}.
\end{equation}
We then have
\begin{equation}
{1 \over z} {d^2 z \over d\tau^2} = 2 a^2 H^2 \left(1 + \epsilon - {3 \over 2}
 \eta + \epsilon^2 - 2 \epsilon \eta + {1 \over 2} \eta^2 + {1 \over 2}
 \xi^2\right),
\end{equation}
where the additional parameters $\eta$ and $\xi^2$ are defined
 as:\cite{copeland93,liddle94}
\begin{equation}
\eta \equiv {m_{\rm Pl}^2 \over 4 \pi} \left({H''\left(\phi\right) \over
 H\left(\phi\right)}\right)\label{eqdefofeta}
\end{equation}
and
\begin{equation}
\xi^2 \equiv {m_{\rm Pl}^4 \over 16 \pi^2} \left({H'\left(\phi\right)
 H'''\left(\phi\right) \over H^2\left(\phi\right)}\right).\label{eqdefofxi}
\end{equation}
These are often referred to as {\em slow roll} parameters, although they are
 defined here without any assumption of slow roll (discussed below). Note that
 despite the somewhat unfortunate standard notation used above, the parameter
 $\xi^2$ can be either positive or negative.

Eq. (\ref{eqexactmodeequation}) can be solved exactly for the case of power-law
 inflation, for which $\epsilon = \eta = \xi = {\rm const.}$, and the scale
 factor evolves as a power-law in time,
\begin{equation}
a\left(t\right) \propto t^{1 / \epsilon} = - {1 \over H \left(1 -
 \epsilon\right)} \tau^{-1}.
\end{equation}
(Note that during inflation, $\tau < 0$, with $\tau \rightarrow 0$ at late
 time.) The vacuum solution to the mode equation (\ref{eqexactmodeequation}) is
 then
\begin{equation}
u_k \propto \sqrt{- k \tau} H_\nu\left(- k \tau\right),\label{eqhankelsolution}
\end{equation}
where $H_\nu$ is a Hankel function of the first kind, and
\begin{equation}
\nu =  {3 \over 2} + {\epsilon \over 1 - \epsilon}.
\end{equation}
The power spectrum for modes with wavelength much larger than the horizon ($k
 \ll a H$) is an exact power law,
\begin{equation}
P_{\cal R}^{1/2} = {H \over 2 \pi \sqrt{\epsilon}}\Bigg\vert_{a H = k} \propto
 k^{n - 1},
\end{equation}
where the spectral index $n$ is given by
\begin{equation}
n = 1 - {2 \epsilon \over 1 - \epsilon}.
\end{equation}
Similarly, the tensor fluctuation amplitude is 
\begin{equation}
P_{\rm T}^{1/2} = {H \over 2 \pi}\Bigg\vert_{a H = k} \propto k^{n_{\rm T}},
\end{equation}
where
\begin{equation}
n_{\rm T} = - {2 \epsilon \over 1 - \epsilon}.
\end{equation}
Other classes of exact solution are known\cite{easther95}.

There are also classes of approximate solution. The standard slow roll
 approximation is the assumption that the field evolution is dominated by drag
 from the expansion, $\ddot\phi \simeq 0$, so that $\dot\phi$ is approximately
 constant and $H\left(\phi\right)$ can be taken to vary as
\begin{equation}
H\left(\phi\right) = \sqrt{\left({8 \pi \over 3 m_{\rm Pl}^2}\right)
 V\left[\phi\left(t\right)\right]},
\end{equation}
where $\phi\left(t\right)$ satisfies
\begin{equation}
\dot\phi = - {V'\left(\phi\right) \over 3 H\left(\phi\right)}.
\end{equation}
This approximation is consistent as long as the first two derivatives of the
 potential are small relative to its magnitude $V',\ V'' \ll V$. The parameters
 $\epsilon$ and $\eta$ reduce in this limit to\cite{kolb94} 
\begin{eqnarray}
&&\epsilon \simeq {m_{\rm Pl}^2 \over 16 \pi} \left({V'\left(\phi\right) \over
 V\left(\phi\right)}\right)^2,\cr
&&\eta \simeq {m_{\rm Pl}^2 \over 8 \pi}\left[{V''\left(\phi\right) \over
 V\left(\phi\right)} - {1 \over 2} \left({V'\left(\phi\right) \over
 V\left(\phi\right)}\right)^2\right].\label{eqssrepsiloneta}
\end{eqnarray}
The slow roll limit can then be equivalently expressed as $\epsilon,\
 \left|\eta\right| \ll 1$. These expressions are frequently taken in the
 literature as definitions of the slow roll parameters, but here they are simply
 limits of the defining expressions (\ref{eqdefofepsilon}) and
 (\ref{eqdefofeta}). In the limit where slow roll is valid, the tensor and
 scalar spectra are again power laws, where the spectral index $n$ is given by
\begin{equation}
n = 1 - 4 \epsilon + 2 \eta.
\end{equation}
The tensor spectral index is just the $\epsilon \ll 1$ limit of the power-law
 case,
\begin{equation}
n_{\rm T} = - 2 \epsilon.
\end{equation}

A second class of approximate models has $\epsilon \ll 1$ as in the slow roll
 case, but is characterized by a large parameter $\eta \simeq {\rm const.} \sim
 O(1)$. In this case, the slow roll expressions (\ref{eqssrepsiloneta}) do not
 apply, and it can be shown that $\epsilon$ can be expressed in terms of the
 potential by\cite{kinney97}
\begin{equation}
\epsilon\left(\phi\right) \simeq 3 \left(1 - {V\left(\phi\right) \over
 V\left(\phi_0\right)}\right),\label{eqnsrepsilon}
\end{equation}
where $\phi_0$ is a stationary point of the field, $V'\left(\phi_0\right) = 0$.
 In this case, the scalar spectral index is 
\begin{equation}
n \simeq 1 + 2 \eta,
\end{equation}
and the tensor spectral index is, as usual,
\begin{equation}
n_{\rm T} \simeq - 2 \epsilon.
\end{equation}
Such models are strongly observationally disfavored, because they predict a
 rapidly varying power spectrum, $\left|n - 1\right| \sim O(1)$, but they are
 nonetheless important as attractors in the inflationary parameter space.

Note that in all cases, the ratio of tensor to scalar perturbations is just the
 first slow roll parameter
\begin{equation}
r \equiv {P_{\rm T} \over P_{\rm R}} = \epsilon.\label{eqdefofr}
\end{equation}
This expression is exact in the power-law case, and valid to lowest order in the
 slow roll case.\footnote{Conventions for the normalization of this parameter
 vary widely in the literature. In particular, the ratio $T/S$ of the tensor and
 scalar contributions to the CMB depends on the current values of $\Omega_{\rm
 M}$ and $\Omega_{\Lambda}$\cite{turner93}. For the currently favored values
 $\Omega_{\rm M} \sim 0.3$, $\Omega_{\lambda} \sim 0.7$, the relationship is
 $T/S \simeq 10 r$, which is the normalization used in Refs. 
\cite{hoffman00,hannestad02}. To compare with the normalization for $r$ as 
defined in Refs. \cite{kinney01,dodelson97,kinney98}, take $r \rightarrow 13.6 
r$.} Note that this is related to the tensor spectral index by
 the inflationary ``consistency condition''\cite{kolb94} $n_{\rm T} = - 2 r$, so
 the shape of the tensor spectrum does not provide an additional independent
 observable. The scalar spectral index depends in general on $\eta$, and
 therefore is an independent observable. (In Section \ref{secfixedpoints} we
 discuss the generalization of the expressions in this section to higher order
 in slow roll.)

\section{The slow roll hierarchy and flow in the inflationary parameter space}
\label{secsrhierarchy}

The slow roll parameters are not in general constant during inflation, but
 change in value as the scalar field driving the inflationary expansion evolves.
 From the definitions (\ref{eqdefofepsilon},\ref{eqdefofeta},\ref{eqdefofxi}) of
 the parameters $\epsilon,\eta,\xi$, we have
\begin{eqnarray}
{d \epsilon \over d \phi} &=& \left({2 \sqrt{\pi} \over m_{\rm Pl}}\right) 2
 \sqrt{\epsilon} \left(\eta - \epsilon\right),\cr
{d \eta \over d \phi} &=&  \left({2 \sqrt{\pi} \over m_{\rm Pl}}\right) {1 \over
 \sqrt{\epsilon}} \left(\xi^2 - \epsilon \eta\right).
\end{eqnarray}
It will be convenient to use the number of e-folds before the end of inflation
 $N$ as the evolution parameter instead of the field. From Eq. (\ref{eqdefofN}),
 it is straightforward to re-write derivatives with respect to $\phi$ in terms
 of derivatives with respect to $N$,
\begin{equation}
{d \over d N} = { m_{\rm Pl} \over 2 \sqrt{\pi}} \sqrt{\epsilon} {d \over
 d\phi}.
\end{equation}
In terms of N, we then have
\begin{equation}
{d \epsilon \over d N} = 2 \epsilon \left(\eta -
 \epsilon\right),\label{eqepsilonflow}
\end{equation}
and
\begin{equation}
{d \eta \over d N} =  \xi^2 - \epsilon \eta.\label{eqetaflow}
\end{equation}
Note that the derivative of each slow roll parameter is itself higher order in
 slow roll. This suggests an infinite hierarchy of ``Hubble slow roll''
 parameters\footnote{The slow roll parameters ${}^\ell\lambda_{\rm
 H}$ used here are related to the parameters ${}^\ell\beta_{\rm H}$ defined by
 Liddle {\it et al.} by ${}^\ell\lambda_{\rm H} = \left({}^\ell\beta_{\rm
 H}\right)^\ell$.}\cite{liddle94}
\begin{equation}
{}^\ell\lambda_{\rm H} \equiv \left({m_{\rm Pl}^2 \over 4 \pi}\right)^\ell
 {\left(H'\right)^{\ell-1} \over H^\ell} {d^{(\ell+1)} H \over d\phi^{(\ell +
 1)}}.\label{eqdefoflambda}
\end{equation}
For example, the $\ell = 2$ parameter is just $\xi^2$:
\begin{equation}
{}^2\lambda_{\rm H} = {m_{\rm Pl}^4 \over 16 \pi^2} \left({H' H''' \over
 H^2}\right) = \xi^2.
\end{equation}
We can then define an infinite hierarchy of ``flow'' equations for the slow roll
 parameters by differentiating Eq. (\ref{eqdefoflambda}),
\begin{equation}
{d \left({}^\ell\lambda_{\rm H}\right) \over d N} = \left[\left(\ell - 1\right)
 \eta - \ell \epsilon\right] \left({}^\ell\lambda_{\rm H}\right) +
 {}^{\ell+1}\lambda_{\rm H}.\label{eqlambdaflow}
\end{equation}
Together with Eqs. (\ref{eqepsilonflow},\ref{eqetaflow}), these form a system of
 differential equations that can be numerically integrated to arbitrarily high
 order in slow roll.\footnote{A similar expansion was derived in Ref. 
\cite{schwarz01}.}

This flow equation approach to studying the inflationary parameter space was
 first suggested by Hoffman and Turner\cite{hoffman00}, who wrote the flow
 equations to lowest order in slow roll in terms of the tensor/scalar ratio
 $(T/S)$  and the scalar spectral index $n$ as:
\begin{eqnarray}
{d\left(T/S\right) \over d N} &=& \left(n - 1\right){T \over S} + {1 \over 5}
 \left({T \over S}\right)^2,\cr
{d\left(n - 1\right) \over d N} &=& -{1 \over 5}\left(n - 1\right) {T \over S} -
 {1 \over 25} \left({T \over S}\right)^2 \pm {m_{\rm Pl}^3 \over 16 \pi^2}
 \sqrt{{2 \pi \over 5} {T \over S}} x'',\label{eqsturnerflow}
\end{eqnarray}
where
\begin{equation}
x\left(\phi\right) \equiv {V'\left(\phi\right) \over V\left(\phi\right)}.
\end{equation}
Here $(T/S)$ is related to the parameter $r$ defined in Eq. (\ref{eqdefofr}) by
 $(T/S) = 10 r$. Hoffman and Turner ``closed'' the flow equations by assuming
 that $x''$ is small and constant. It is straightforward to generalize these
 equations using the Hubble slow roll formalism above. We can define a new
 parameter 
\begin{equation}
\sigma \equiv 2 \eta - 4 \epsilon,
\end{equation}
which is equivalent the spectral index parameter used by Hoffman and Turner:
 $\sigma \simeq n - 1$ to lowest order in slow roll. The flow equations
 (\ref{eqepsilonflow},\ref{eqetaflow}) in terms of $\sigma$ are:
\begin{eqnarray}
{d \epsilon \over d N} &=& \epsilon \left(\sigma + 2 \epsilon\right),\cr
{d \sigma \over d N} &=& 2 \xi^2 - 5 \epsilon \sigma - 12
 \epsilon^2.\label{eqsigmaflow}
\end{eqnarray}
These expressions can be shown to be identical to Eqs. (\ref{eqsturnerflow}) by
 evaluating using the slow roll expressions (\ref{eqssrepsiloneta}) for
 $\epsilon$ and $\eta$. Using $\xi^2 = {}^2\lambda_{\rm H}$, the flow equations
 (\ref{eqsigmaflow}) along with (\ref{eqlambdaflow}) then represent a
 generalization of the flow equations of Hoffman and Turner to arbitrarily high
 order in slow roll. This system of equations, taken to infinite order, is
 exact. In practice, these equations must be truncated at some finite order, by
 assuming ${}^\ell\lambda_{\rm H} = 0$ for $\ell$ greater than some finite order
 $M$. The higher order the truncation, the weaker the implicit assumptions about
 the form of the potential. The next section discusses the fixed point structure
of the inflationary parameter  space.

\section{Fixed points in the inflationary parameter space}
\label{secfixedpoints}

Summarizing the results of the previous section, the hierarchy of inflationary
 flow equations is:
\begin{eqnarray}
{d \epsilon \over d N} &=& \epsilon \left(\sigma + 2 \epsilon\right),\cr
{d \sigma \over d N} &=& - 5 \epsilon \sigma - 12 \epsilon^2 + 2
 \left({}^2\lambda_{\rm H}\right),\cr
{d \left({}^\ell\lambda_{\rm H}\right) \over d N} &=& \left[{1 \over 2}
 \left(\ell - 1\right) \sigma + \left(\ell - 2\right) \epsilon\right]
 \left({}^\ell\lambda_{\rm H}\right) + {}^{\ell+1}\lambda_{\rm
 H}.\label{eqfullflowequations}
\end{eqnarray}
To lowest order in slow roll, these can be related to observables by $r =
 \epsilon$ and $n - 1 = \sigma$. To second order in slow roll, the observables
 are given by\cite{stewart93}
\begin{equation}
r = \epsilon \left[1 - C \left(\sigma + 2
 \epsilon\right)\right],\label{eqrsecondorder}
\end{equation}
for the tensor/scalar ratio, and 
\begin{equation}
n - 1 = \sigma - \left(5 - 3 C\right) \epsilon^2 - {1 \over 4} \left(3 - 5
 C\right) \sigma \epsilon + {1 \over 2}\left(3 - C\right) \left({}^2\lambda_{\rm
 H}\right)\label{eqnsecondorder}
\end{equation}
for the spectral index. Here $C \equiv 4 \left(\ln{2} + \gamma\right) - 5 = 
0.0814514$, 
where $\gamma \simeq 0.577$ is Euler's constant.\footnote{Earlier versions of 
this paper
 followed Ref. \cite{liddle94}, which incorrectly specifies 
 $C \equiv 4 \left(\ln{2} + \gamma\right) = 5.0814514$. This is apparently a 
typographic
 error in Ref. \cite{liddle94}.}
Derivatives with respect to
wavenumber $k$ can be expressed in terms of derivatives with respect to $N$ 
as\cite{liddle95}
\begin{equation}
{d \over d N} = - \left(1 - \epsilon\right) {d \over d \ln k},
\end{equation}
The scale dependence of $n$ is then  given by the simple expression
\begin{equation}
{d n \over d \ln k} = - \left({1 \over 1 - \epsilon}\right) {d n \over d N},
\end{equation}
which can be evaluated to third order in slow roll by using Eq.
 (\ref{eqnsecondorder}) and the flow equations. We wish to study flow in the
parameter space of  observables, $r$, $n$, and $d n / d \ln k$.

It is useful to identify fixed points of the system of equations
 (\ref{eqfullflowequations}), for which all the derivatives vanish. Two classes
 of fixed points are easily obtained by inspection. First is the case of
 vanishing tensor/scalar ratio, with
\begin{eqnarray}
\epsilon &=& {}^\ell\lambda_{\rm H} = 0,\cr
\sigma &=& {\rm const.}\label{eqfixedpoint1}
\end{eqnarray}
The second class of fixed points is just the case of power-law inflation,
 $\epsilon = \eta = \xi^2 = {\rm const.}$, or
\begin{eqnarray}
\epsilon &=& {\rm const.},\cr
\sigma &=& - 2 \epsilon,\cr
{}^2\lambda_{\rm H} &=& \epsilon^2,\cr
{}^{\ell+1}\lambda_{\rm H} &=& \epsilon \left({}^\ell\lambda_{\rm H}\right),\
 \ell \geq 2.\label{eqfixedpoint2}
\end{eqnarray}
Note that these are fixed points of the exact system of equations. It is
 straightforward to evaluate the stability of the fixed point
 (\ref{eqfixedpoint1}), since
\begin{equation}
{d^2 \epsilon \over dN d\epsilon}\Bigg\vert_{\epsilon = 0} = \sigma,
\end{equation}
and
\begin{equation}
{d^2 \epsilon \over d N d\sigma}\Bigg\vert_{\epsilon = 0} = 0.
\end{equation}
Therefore the fixed point at $\epsilon = 0$ is stable with respect to
 perturbations in $\epsilon$ for $\sigma > 0$, or spectral index $n > 1$, and
 unstable for $\sigma < 0$, or spectral index $n < 1$. (This unusual sign
 convention for stability comes from the definition $d N < 0$ for $dt > 0$.) In
 general, inflationary evolution flows away from $r = 0$ for $n < 1$, and toward
 $r = 0$ for $n > 1$. This behavior can be easily understood in terms of simple
 inflaton potentials in slow roll. Using the slow roll expressions
 (\ref{eqssrepsiloneta}), taking $\epsilon = 0$ implies that the field is at an
 equilibrium point $\dot \phi \propto V'\left(\phi\right) = 0$, and the spectral
 index is 
\begin{equation}
n - 1 = 2 \eta \simeq {m_{\rm Pl}^2 \over 4 \pi} {V''\left(\phi\right) \over
 V\left(\phi\right)}.
\end{equation}
The case $\epsilon = 0$, $n < 1$ is just that of the field sitting atop an
 unstable equilibrium, for example the point $\phi = 0$ on a potential of the
 form $V\left(\phi\right) = \Lambda^4 - m^2 \phi^2$. The case $\epsilon = 0$, $n
 > 1$ is that of a field sitting at a stable equilibrium point $V'' > 0$, for
 example the point $\phi = 0$ on a potential of the form $V\left(\phi\right) =
 \Lambda^4 + m^2 \phi^2$. In such models, inflation nominally continues forever.
 In practice, however, it is possible to end inflation by coupling to additional
 fields, as in ``hybrid'' inflation models\cite{linde91,linde94,copeland94}. The
 observables in this a case are given by their values near the late-time
 asymptote. The case of the fixed point (\ref{eqfixedpoint2}) is more complex. 
It
 is, however, known that it is not in general a late-time
 attractor\cite{escalante02}, a conclusion that is supported by numerical
 integration of the flow equations.

\section{Evaluating the flow equations}
\label{secevaluatingtheflowequations}

With the flow equations in hand, it is possible to ask the question: what are
 the {\em generic} predictions of inflation? In principle, any model of
 inflation driven by a single, monotonic scalar field can be completely
 specified by selecting a point in the (infinite dimensional) slow roll
 parameter space, $\epsilon,\sigma,{}^\ell\lambda_{\rm H}$.\footnote{Strictly
 speaking, this statement is true only if the slow roll expansion is nonsingular
 to all orders.} For a model specified in this way, there is a straightforward
 procedure for determining its observable predictions, that is, the values of
 $r$, $n - 1$, and $d n / d \ln k$ a fixed number $N$ e-folds before the end of
 inflation. The algorithm for a single model is as follows:
\begin{itemize}
\item{Select a point in the parameter space $\epsilon,\eta,{}^l\lambda_{\rm
 H}$.}
\item{Evolve forward in time ($d N < 0$) until either (a) inflation ends, or (b)
 the evolution reaches a late-time fixed point.}
\item{If the evolution reaches a late-time fixed point, calculate the
 observables $r$, $n - 1$, and $d n / d \ln k$ at this point.}
\item{If inflation ends, evaluate the flow equations backward $N$ e-folds from
 the end of inflation. Calculate the observable parameters at this point.}
\end{itemize}
The end of inflation is given by the condition $\epsilon = 1$ ({\em not} by the
 end of slow roll, although in practice these conditions are essentially
 equivalent). In the case where inflation ends in the late-time limit, there is
 another possibility: that one will find that inflation also ends when evolving
 back to early times. That is, the model is incapable of supporting $N$ e-folds
 of inflation.

In principle, it is possible to carry out this program exactly, with no
 assumptions made about the convergence of the hierarchy of slow roll
 parameters. In practice, the series of flow equations
 (\ref{eqfullflowequations}) must be truncated at some finite order and
 evaluated numerically. In addition, for any given path in the parameter space,
 we do not know {\it a priori} the correct number of e-folds $N$ at which to
 evaluate the observables, since this depends on details such as the energy
 density during inflation and the reheat temperature\cite{lidsey95}. We adopt a
 Monte Carlo approach: we evaluate a large number of inflation models at order
 $M$ in slow roll, where each model consists of a randomly selected set of
 parameters in the following ranges:
\begin{eqnarray}
N &=& [40,70]\cr
\epsilon &=& \left[0,0.8\right]\cr
\sigma &=& \left[-0.5,0.5\right]\cr
{}^2\lambda_{\rm H} &=& \left[-0.05,0.05\right]\cr
{}^3\lambda_{\rm H} &=& \left[-0.005,0.005\right],\cr
&\cdots&\cr
{}^{M+1}\lambda_{\rm H} &=& 0.\label{eqinitialconditions}
\end{eqnarray}
and so forth, reducing the width of the range by factor of ten for each higher
 order in slow roll. The series is closed to order $M$ by taking
 ${}^{M+1}\lambda_{\rm H} = 0$. The exact choice of ranges for the initial
 parameters does not have a large influence on the result of the Monte Carlo, as
 long as they are chosen such that the slow roll hierarchy is convergent. For
 each model, we calculate observables according to the algorithm above, with two
 differences because of the finite nature of the calculation. When we evolve
 forward in time, there are now three possible late-time behaviors for a
 particular model: (1) the model reaches the late-time attractor $\epsilon = 0$,
 $\sigma > 0$, (2) inflation ends, or (3) none of the above, indicating that the
 integration failed to reach any identifiable asymptotic behavior within the
 limits of the integration, which we take to be 1000 e-folds. For models in
 which inflation ends at late time, we then evolve the model backward in time
 $N$ e-folds from the end of inflation. If the choice of parameters supports $N$
 e-folds without inflation ending or slow roll failing, we calculate observable
 parameters $r$, $n$, and $d n / d \ln k$ at that point. We will call these
 points {\em nontrivial} points.
In summary, there are four categories of outcome for a particular choice of
 initial condition
\begin{itemize}
\item{Late-time attractor, $\epsilon = 0$, $\sigma > 0$.}
\item{Insufficient inflation.}
\item{Nontrivial point: Inflation ends at late time, supports N e-folds of
 inflation.}
\item{No identifiable asymptotic behavior at late time.}
\end{itemize}
The numerical integration is implemented in C using a fifth-order adaptive
 step-size Runge-Kutta method to solve the system of equations. The Monte Carlo
 is run by selecting initial conditions at random as described above for 
1,000,000
 points. We are interested in the models which converge to a late-time attractor
 or possess a nontrivial point. In addition, we require $n < 1.5$
in order to be consistent with observations of the Cosmic Microwave
 Background\cite{kinney01,hannestad02} and constraints from primordial black
 hole formation\cite{carr94,bullock96,green97,kotok98}. The results of a Monte
 Carlo run to
 order $M = 5$ in slow roll are as follows:
\begin{itemize}
\item{Total iterations: 1,000,000.}
\item{Late-time attractor, $r = 0$, $n > 1.5$: 902,407.}
\item{Nontrivial points: 78,930.}
\item{Late-time attractor, $r = 0$, $n < 1.5$: 16,759.}
\item{Insufficient inflation: 1899.}
\item{No identifiable asymptotic behavior: 5.}
\end{itemize}
One surprising result is that more than 90\% of the models evaluated result in
an  unacceptably blue spectral index, $n > 1.5$: the most ``generic'' prediction
of  inflation from this point of view is already ruled out! Figure 1 shows the
 remaining models plotted on the $(n,r)$ plane. (Note that the normalization for 
$r$ used here differs from elsewhere in the literature. To compare with Refs. 
\cite{kinney01,dodelson97,kinney98}, take $r \rightarrow 13.6 r$. To compare 
with Refs. \cite{hoffman00,hannestad02}, take $r \rightarrow T/S = 10 r$.) The 
models cluster strongly near
 (but {\em not} on) the power-law fixed point, and on the $r = 0$ fixed point.
 This is qualitatively consistent with the results of Hoffman and Turner, except
 that the models appear to be much more strongly clustered in the parameter
 space than they concluded from a lowest-order analysis. Also, models sparsely
 populate the regions that Hoffman and Turner label ``excluded'' and ``poor
 power law'', suggesting that these categorizations do not generalize to higher
 order in slow roll. (We note that a poor power law was a
 rare result in the integrations, with of order 0.1\% of the models
predicting  $\left\vert dn/d\ln k\right\vert > 0.05$.)  Figure 2 shows the 
$(n,r)$ plane zoomed in on 
the observationally favored region near $n = 1$. Figure 3 shows the same models 
plotted vs. $\log(r)$ to show the small-$r$ behavior of the attractor region. 
Figure 4 shows  the same
models plotted on the $n, dn/d\ln k$ plane, also showing noticeable  clustering
behavior in the parameter space. In particular, $d n / d \ln k < 0$  is favored.

\begin{figure}
\centerline{\psfig{figure=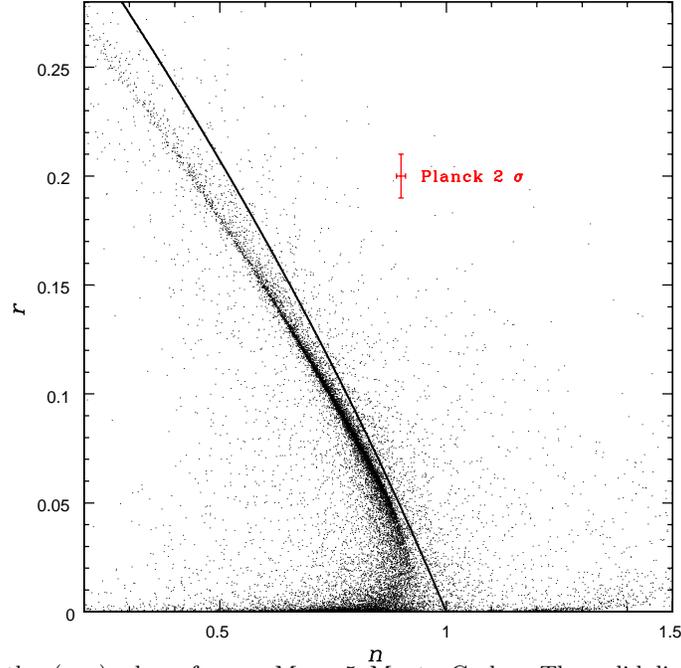,width=3.5in}} 
\caption{Models plotted in the $(n,r)$ plane for an $M=5$ Monte Carlo. The solid
line is the power-law fixed point $n = 1 - 2 r / (1 - r)$. The error bar
shows the size of the expected $2\ \sigma$ error from Planck. (See note in text 
regarding the somewhat unconventional normalization of $r$ used here.)}
\end{figure}
\begin{figure}
\centerline{\psfig{figure=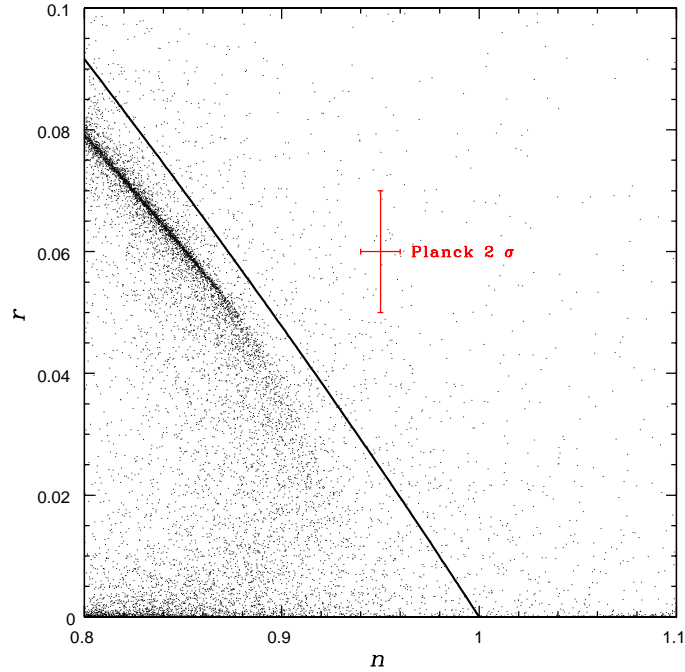,width=3.5in}} 
\caption{Figure 1 zoomed in to the region preferred by observation.}
\end{figure}
\begin{figure}
\centerline{\psfig{figure=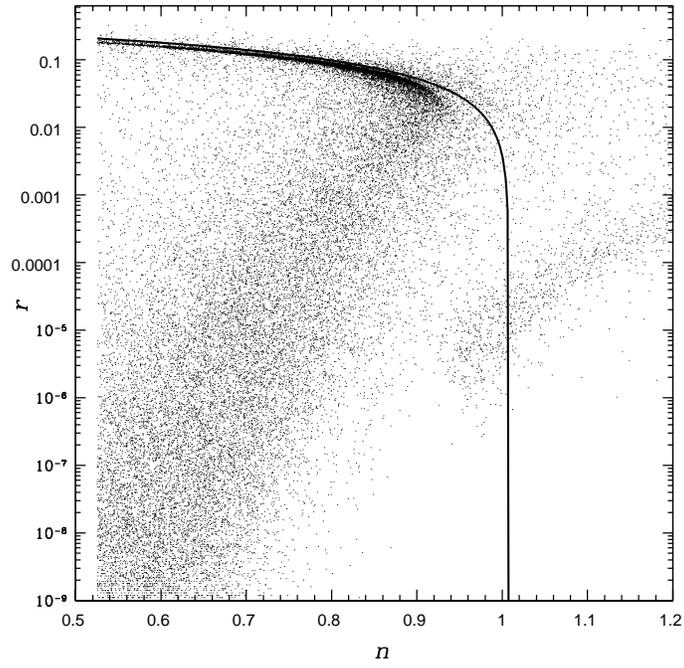,width=3.5in}} 
\caption{Spectral index vs. $\log(r)$, showing the behavior of the attractor 
region for small r.}
\end{figure}
\begin{figure}
\centerline{\psfig{figure=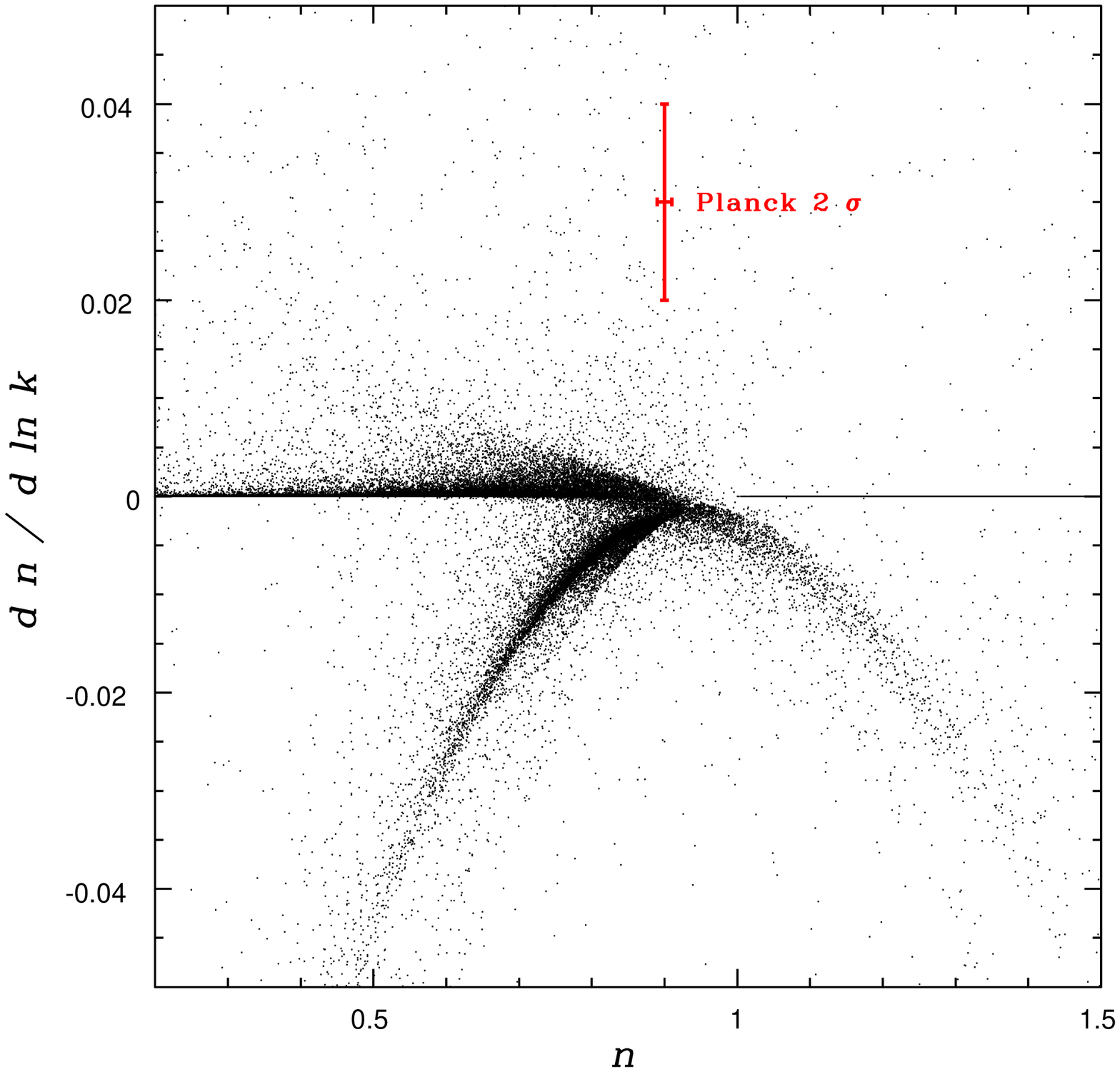,width=3.5in}} 
\caption{Models plotted in the $(n,dn/d\ln k)$ plane for an $M=5$ Monte Carlo.}
\end{figure}
\begin{figure}
\centerline{\psfig{figure=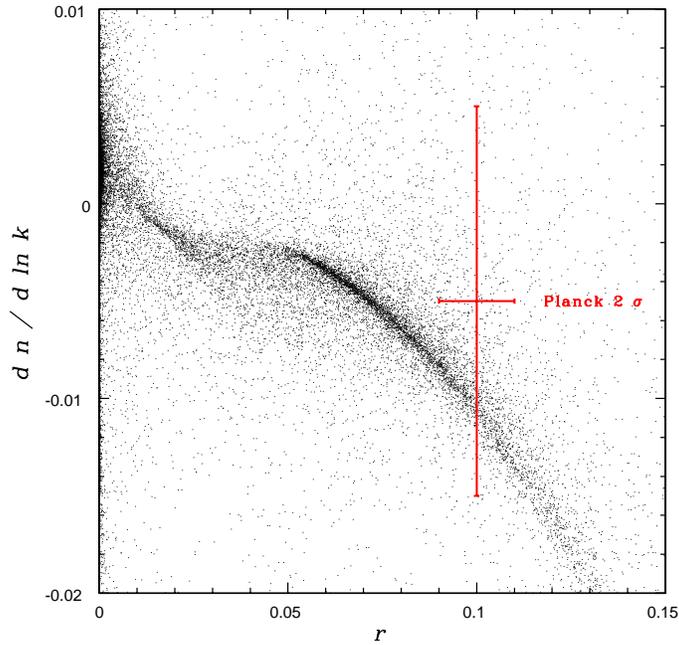,width=3.5in}} 
\caption{Models plotted in the $(r,dn/d\ln k)$ plane for an $M=5$ Monte Carlo.}
\end{figure}

Figure 5 shows $d n / d \ln k$ as a function of $r$. Especially  interesting is
that the models with large $r$ (the ones close to the power-law  line in Fig.
1), also have significant variation in the spectral index. This  suggests that
the models are not flowing to the power-law fixed point, which  has $d n / d \ln
k = 0$. This raises an interesting question: are the models  converging slowly
to the power-law line at early times, or are they converging  to some other
fixed point? To answer this question, we evolve the models to  very early times,
$N >> 70$. Figure 6 shows models plotted on the $(n,r)$ plane  for $N = 125,\
250,\ 500$, and $1000$. Instead of flowing to the power-law  fixed point at
early time, the models instead flow down to the $r = 0$ line. We  therefore find
that the power-law fixed point is not an attractor at early or  late time. It is
important to note that this conclusion is not in conflict with  the analysis of
Copeland {\it et al.}\cite{copeland98}, which concluded that  power-law
inflation is a unique late-time attractor in a cosmology consisting  of a scalar
field and a second fluid component. Copeland {\it et al.} assumed a  scalar
field with an exponential potential -- that is, a model lying exactly on  the
power-law fixed point -- and showed that the scalar field would generically
dominate the cosmological evolution at late times. Figure 7 shows examples of
flow in the $(\sigma,\epsilon)$ plane.

\begin{figure}
\centerline{\psfig{figure=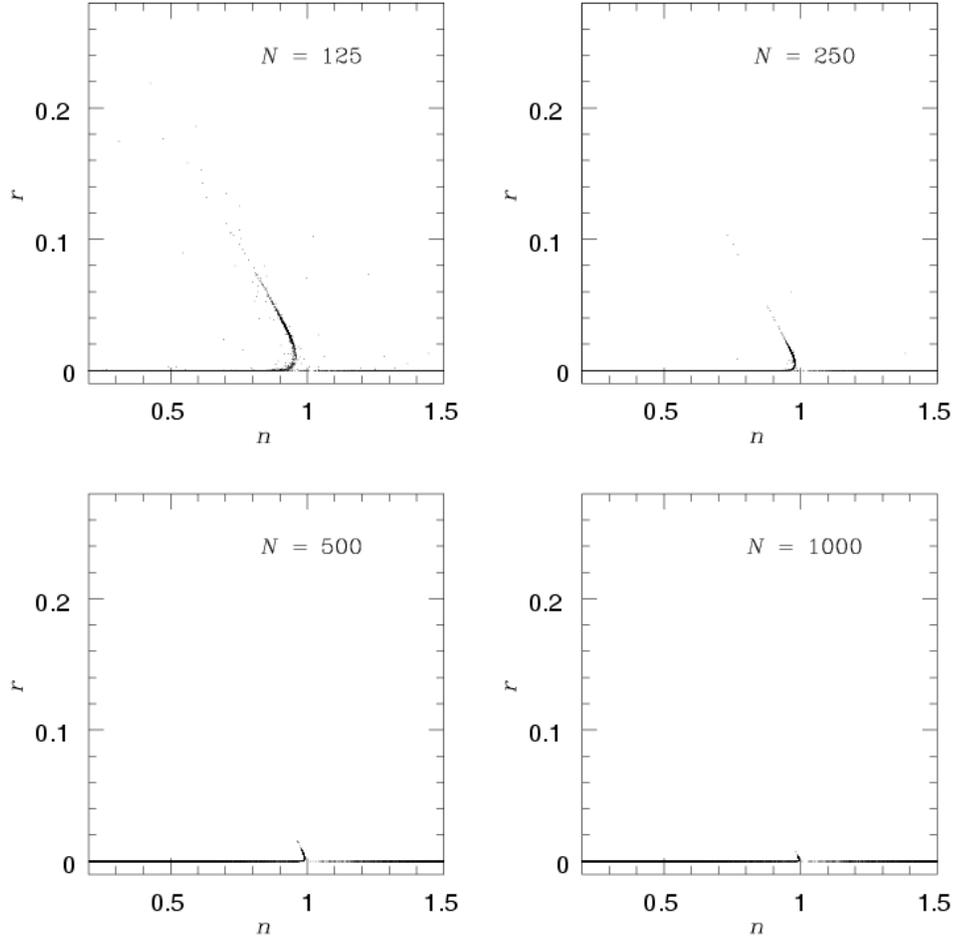,width=5.5in}} 
\caption{Results of the Monte Carlo with a large number of e-folds $N$,
 illustrating the behavior of the flow at early times (larger $N$). The models
 flow not to the power-law fixed point $n = 1 - 2 r /(1 - r)$ but to the $r = 0$
 fixed point.}
\end{figure}
\begin{figure}
\centerline{\psfig{figure=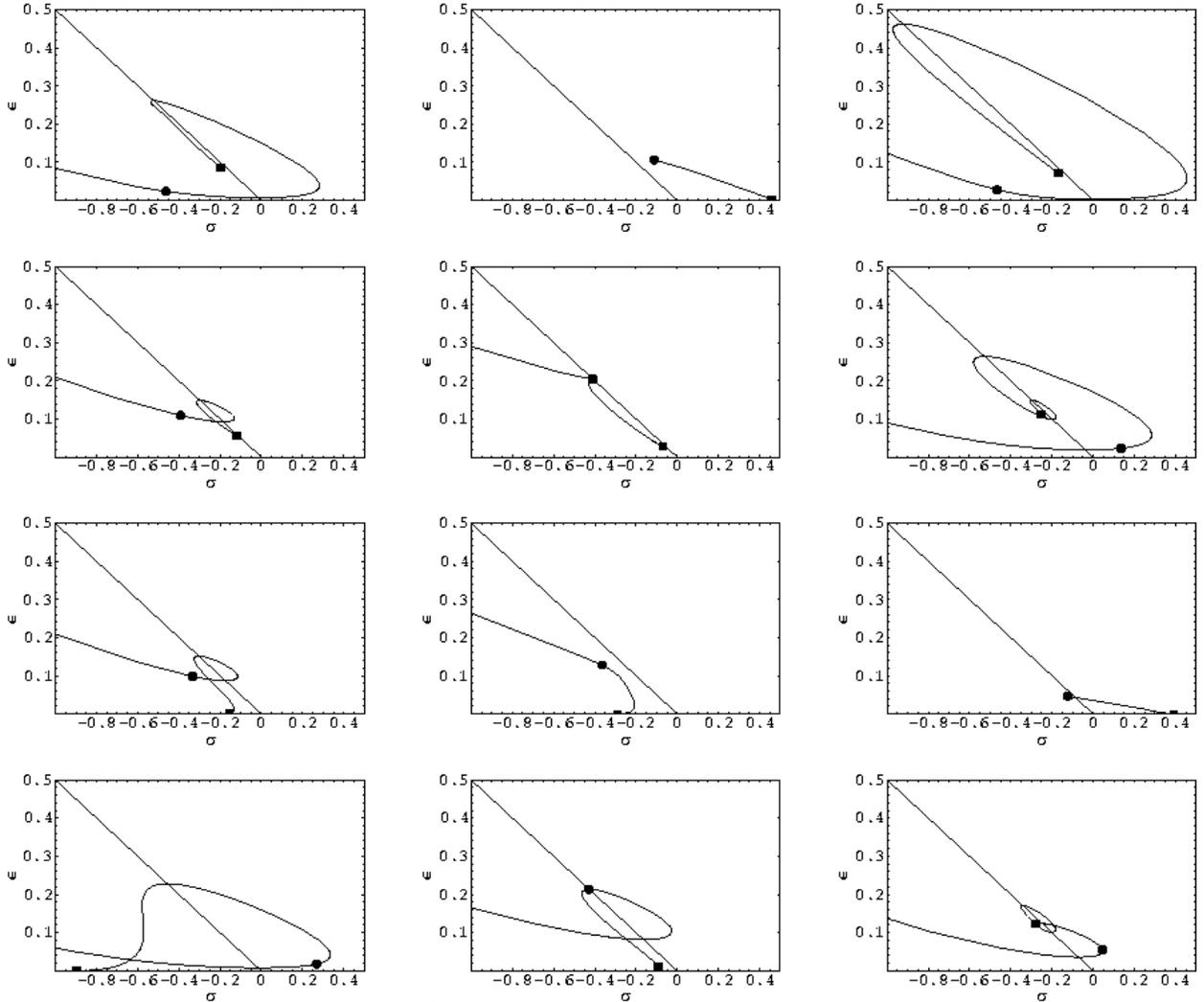,width=6.5in}} 
\caption{Examples of flow plotted in the $(\sigma,\epsilon)$ plane. The circles
 indicate the randomly selected initial value, and the squares indicate the
 value $N$ e-folds before the end of inflation.  The straight line is the
 power-law fixed point, $\sigma = -2 \epsilon$. Integrating to high order in
 slow roll allows for a variety of complex flows.}
\end{figure}

\section{Conclusions}
\label{secconclusions}

We have derived a set of inflationary ``flow'' equations based on the Hubble
slow roll expansion of Liddle {\it et al.}\cite{liddle94} that is in principle
exact when taken to all orders. These equations completely specify the dynamics
of the inflationary system, so that any particular inflationary potential can
be specified as a point in this parameter space. The past and future dynamics
of the model are then determined by evaluating the flow of the parameters away
from this point. It is possible to identify two classes of fixed points of the
exact flow equations: power-law inflation, with $n = 1 - 2 r / (1 - r)$, and
models with vanishing tensor/scalar ratio, $r = 0$. This latter class is
unstable for $n < 1$ and stable for $n > 1$.

In practice, the flow equations must be truncated to some order and evaluated
numerically, which was done to lowest order by Hoffman and
Turner\cite{hoffman00}. Extending the system of flow equations to higher order
makes it possible to consider the running of the spectral index $d n / d \ln k$
as well as $r$ and $n$. We perform a Monte Carlo integration of the flow
equations to fifth order in slow roll, and show that the distribution of models
in the parameter space of observables $r$, $n$ and $d n /d \ln k$ is strongly
clustered around particular values. Ninety percent of the models selected in the
Monte Carlo converge to the observationally unacceptable asymptote $r = 0$, $n >
1.5$. The remaining models cluster around two classes of early-time
``attractor'', the first class at the $r = 0$ fixed point and the second with $r
> 0$ and $n < 1$. Interestingly, the $r > 0$ attractor {\em cannot} be
identified with the power-law fixed point, since they generally have $d n / d
\ln k < 0$, and the variation in the spectral index vanishes at the fixed point.
Evaluation of the models at very early times, $N \gg 70$, indicates that the
power-law fixed point is not an attractor at early times, since the models
generically flow to the $r = 0$ line for large $N$. We therefore interpret the
$r > 0$ ``attractor'' as simply an artifact of the fact that observable
perturbations are generated relatively late in the inflationary evolution, when
slow roll has begun to measurably break down. In addition, we see that power-law
inflation is not in general an attractor for either early {\em or} late times.
At higher order, models much cluster more strongly than is suggested by the
``favored'' region of the parameter space derived by Hoffman and Turner. Also,
models sparsely populate the regions labeled by Hoffman and Turner as 
``excluded'' and ``poor power law'', suggesting that these categorizations do
not generalize to higher order. 

It is important to consider questions of generality with respect to both the
choice of the order in slow roll $M$ and the choice of initial conditions for
the Monte Carlo (\ref{eqinitialconditions}). By ``closing'' the hierarchy of
flow equations at finite order, we are implicitly limiting ourselves to a
restricted class of potentials, although for $M = 5$, that class of potentials
is large. However, models with potentials that contain
features\cite{starobinsky92,adams01} or for which the slow roll expansion is
not convergent\cite{wang97} will not be captured by solutions at finite order
in slow roll. In addition, inflation might not be driven by only a single
scalar field. The effect of different choices of initial conditions can be
studied empirically, simply by trying different constraints on the space of
initial conditions. Choosing ``looser'' initial conditions does not alter the
characteristics of the result. Instead, models which fail to support sufficient
inflation become much more numerous. Perhaps most importantly, absent a
metric on the space of initial conditions, one should use caution when
attempting to interpret these ``scatter plots'' statistically. We do not
know how the initial conditions for the universe were selected! However, if
observations determine that the relevant cosmological parameters lie outside the
``favored'' region, it will be an indication of highly unusual dynamics during
the inflationary epoch.

Finally, we note an interesting recent body of literature connecting flow in
inflationary models to a proposed ``holographic'' correspondence between
quasi-de Sitter spaces and boundary conformal field theories
(CFTs)\cite{strominger01a,strominger01b,klemm01,bala01,bousso01,medved01,halyo02}.
In particular, Larsen {et al.} have proposed a correspondence between slow
roll parameters and couplings in the boundary CFT, interpreting flow in the
inflationary parameter space as renormalization group flow in the associated
CFT\cite{larsen02,argurio02}. The fixed points at $r = 0$ are interpreted as
ultraviolet ($n > 1$) and infrared ($n < 1$) fixed points in the renormalization
group flow. In this picture, studying inflationary dynamics is equivalent to
studying the structure of the underlying CFT. (It is not immediately clear,
however, how one interprets the power-law fixed point in the context of the
boundary CFT.)

\section*{Acknowledgments}

WHK is supported by ISCAP and  the Columbia University Academic Quality Fund.
ISCAP gratefully acknowledges the generous support of the Ohrstrom Foundation.
I would like to thank Richard Easther, Rocky Kolb, Andrew Liddle, Koenraad Schalm
and Jan-Pieter van der Schaar for helpful conversations.


\begin{references}
\bibitem{guth81} A. H. Guth, Phys. Rev. D {\bf 23}, 347 (1981).
\bibitem{linde82} A. D. Linde, Phys. Lett. {\bf 108B}, 389 (1982).
\bibitem{albrecht82} A. Albrecht and P. J. Steinhardt, Phys. Rev. Lett. {\bf
 48}, 1220 (1982).
\bibitem{lyth99} D.~H.~Lyth and A.~Riotto, Phys. Rep. {\bf 314}, 1 (1999),
 hep-ph/9807278.
\bibitem{kinney01} W.~H.~Kinney, A.~Melchiorri, and A.~Riotto, Phys. Rev. D {\bf
 63}, 023505  (2001), astro-ph/0007375.
\bibitem{hannestad02} S.~Hannestad, S.~H.~Hansen, and F.~L.~Villante, Astropart.
 Phys. {\bf 17} 375 (2002),  astro-ph/0103047.
\bibitem{MAP} http://map.gsfc.nasa.gov/
\bibitem{Planck} http://astro.estec.esa.nl/SA-general/Projects/Planck/
\bibitem{dodelson97} S.~Dodelson, W.~H.~Kinney, and E.~W.~Kolb,  Phys. Rev. D
 {\bf 56}, 3207 (1997), astro-ph/9702166.
\bibitem{kinney98} W.~H.~Kinney,  Phys. Rev. D {\bf 58}, 123506 (1998),
 astro-ph/9806259.
\bibitem{hoffman00} M.~B.~Hoffman and M.~S.~Turner, Phys. Rev. D {\bf 64},
 023506 (2001), astro-ph/0006321.
\bibitem{strominger01a} A.~Strominger, hep-th/0110087.
\bibitem{larsen02}   F.~Larsen, J.~P.~van~der~Schaar, and R.~G.~Leigh,
 hep-th/0202127.
\bibitem{grishchuk88} L. P. Grishchuk and Yu. V. Sidorav, in {\it Fourth Seminar
 on Quantum Gravity}, eds M. A. Markov, V. A. Berezin and V. P. Frolov (World
 Scientific, Singapore, 1988).
\bibitem{muslimov90} A. G. Muslimov, Class. Quant. Grav. {\bf 7}, 231 (1990).
\bibitem{salopek90} D. S. Salopek and J. R. Bond, Phys. Rev. D {\bf 42}, 3936
 (1990).
\bibitem{lidsey95} J. E. Lidsey {\it et al.}, Rev. Mod. Phys. {\bf 69}, 373
 (1997), astro-ph/9508078.
\bibitem{hawking82} S. W. Hawking, Phys. Lett. {\bf 115B}, 295 (1982).
\bibitem{starobinsky82} A. Starobinsky, Phys. Lett. {\bf 117B}, 175 (1982).
\bibitem{guth82} A.~H.~Guth and S. Y. Pi, Phys. Rev. Lett. {\bf 49}, 1110 
(1982).
\bibitem{bardeen83} J. M. Bardeen, P. J. Steinhardt, and M. S. Turner, Phys.
 Rev. D {\bf 28}, 679 (1983).
\bibitem{mukhanov92} V. F. Mukhanov, H. A. Feldman, and R. H. Brandenberger,
 Phys. Rep. {\bf 215}, 203 (1992).
\bibitem{stewart93} E. D. Stewart and D. H. Lyth, Phys. Lett. {\bf 302B}, 171
 (1993), gr-qc/9302019.
\bibitem{mukhanov85} V. F. Mukhanov, JETP Lett. {\bf 41}, 493 (1985).
\bibitem{mukhanov88} V. F. Mukhanov, Sov. Phys. JETP {\bf 67}, 1297 (1988).
\bibitem{copeland93} E. J. Copeland, E. W. Kolb, A. R. Liddle, and J. E. Lidsey,
 Phys. Rev. D {\bf 48}, 2529 (1993), hep-ph/9303288.
\bibitem{liddle94} A. R. Liddle, P. Parsons, and J. D. Barrow, Phys. Rev. D {\bf
 50}, 7222 (1994), astro-ph/9408015.
\bibitem{easther95} R.~Easther, Class. Quant. Grav. {\bf 13}, 1775 (1996),
 astro-ph/9511143.
\bibitem{kolb94} E. W. Kolb and S. L. Vadas, Phys. Rev. D {\bf 50}, 2479 (1994),
 astro-ph/9403001.
\bibitem{kinney97} W.~H.~Kinney, Phys. Rev. D {\bf 56}, 2002 (1997),
 hep-ph/9702427.
\bibitem{turner93} M.~S.~Turner, M.~White, and J.~E.~Lidsey, Phys. Rev. D {\bf
 48}, 4613 (1993), astro-ph/9306029.
\bibitem{schwarz01} D.~J.~Schwarz, C.~A.~Terrero-Escalante, and A.~.A.~Garcia, 
Phys. Lett. {\bf B517}, 243 (2001), astro-ph/0106020.
\bibitem{liddle95} A.~R.~Liddle. and M.~S.~Turner, Phys. Rev. D {\bf 50}, 758 
(1994). 
\bibitem{linde91} A. D. Linde, Phys. Lett. {\bf 259B}, 38 (1991).
\bibitem{linde94} A. Linde, Phys. Rev. D {\bf 49} 748 (1994),
 astro-ph/9307002.
\bibitem{copeland94} E. J. Copeland {\it et al.}, Phys. Rev. D {\bf 49}, 6410
 (1994),astro-ph/9401011.
\bibitem{escalante02}  C.~A.~Terro-Escalante, astro-ph/0204066.
\bibitem{carr94} B.~J.~Carr, J.~H.~Gilbert, and J.~E.~Lidsey, Phys. Rev. D {\bf
 50}, 4853 (1994), astro-ph/9405027.
\bibitem{bullock96} J.~S.~Bullock and J.~R.~Primack, Phys. Rev. D {\bf 55}, 7423
 (1997), astro-ph/9611106.
\bibitem{green97} A.~M.~Green and A.~R.~Liddle, Phys. Rev. D {\bf 56}, 6166 
(1997),
 astro-ph/9704251.
\bibitem{kotok98} E.~Kotok and P.~Naselsky, Phys. Rev. D {\bf 58}, 103517 
(1998),
 astro-ph/9806139.
\bibitem{copeland98} E.~J.~Copeland, A.~R.~Liddle, and D.~Wands, Phys .Rev. D
 {\bf 57}, 4686 (1998), gr-qc/9711068.
\bibitem{starobinsky92} A.~A.~Starobinsky, JETP Lett. {\bf 55}, 489 (1992).
\bibitem{adams01} J.~Adams, B.~Cresswell, and R.~Easther, Phys. Rev. D {\bf 64},
 123514 (2001), astro-ph/0102236.
\bibitem{wang97} L.~Wang, V.~F.~Mukhanov, and P.~J.~Steinhardt,  Phys. Lett.
 {\bf B414}, 18 (1997).
\bibitem{strominger01b} A.~Strominger, JHEP {\bf 0110}, 034 (2001),
 hep-th/0106113.
\bibitem{klemm01} D.~Klemm,  Nucl. Phys. {\bf B625}, 295 (2002), hep-th/0106247.
\bibitem{bala01} V.~Balasubramanian, J.~de~Boer, and D.~Minic, hep-th/0110108.
\bibitem{bousso01} R.~Bousso, A.~Maloney, and A.~Strominger, Phys. Rev. D {\bf
 65}, 104039 (2002),  hep-th/0112218.
\bibitem{medved01} A.~J.~M.~Medved,  hep-th/0112226.
\bibitem{halyo02} E.~Halyo, hep-th/0203235.
\bibitem{argurio02} R.~Argurrio, hep-th/0202183.
\end{references}
\end{document}